# Comment on the shape of Hydrogen equation in spaces of arbitrary dimension[1]
## (Bohr's molecular model and the melding of classical and quantum mechanics)

**M. Ya. Amusia**

*The Racah Institute of Physics, the Hebrew University of Jerusalem, Jerusalem 91904, Israel*
and
*A. F. Ioffe Physical-Technical Institute, St. Petersburg 194021, Russian Federation*

**Abstract:**
We note that presenting Hydrogen atom Schrodinger equation in the case of arbitrary dimensions require simultaneous modification of the Coulomb potential that only in three dimensions has the form $Z/r$. This was not done in a number of relatively recent papers [1-5]. Therefore some results obtained there seem to be doubtful. Some required considerations in the area are mentioned.

With considerable interest I looked into the article 'Bohr's molecular model, a century later" by Anatoly Svidzinsky, Marlan Scully, and Dudley Herschbach, published in the January 2014 issue of Physics Today. It is always interesting to see how our current understanding sheds new light upon the revolutionary scientific ideas of the past. I agree with the authors that although quantum mechanics is the real bases for atomic and molecular physics computations, the old Bohr's model that treats electrons in atoms like tiny planets moving around the sun-nucleus is intuitively clear and very attractive.

As far as I see it, the aim of the article is to show how to treat using old Bohr's approach not only simple atoms but molecules. An important point in this development is reconciliation of quantum mechanics with Bohr's ideas. The authors claim that in the infinite dimension quantum mechanics "morphs into classical mechanics".

I cannot say, however, that the infinite-dimension system is a clarifying model to describe physical or chemical objects. I do not see that the reference to chromodynamics (and to Edward Witten's Physics Today paper from July 1980) is a clarifying one. Far from being convincing are statements like "Hence the large-D limit, where 1/D→0 is closer to the real world (1/D=1/3) than is the oft-used D=1 regime. Indeed, results obtained at large D usually resemble those for D=3". I must confess, it sounds too light weighted to be convincing.

But not the large-D limit per se bothers me. Of great concern is the assumption that the radial part of the D-dimensional Schrödinger equation in Hartree units looks as follows:

$$\left\{-\frac{1}{2}\frac{\partial^2}{\partial r^2} + \frac{[l+(D-3)/2][l+(D-1)/2]}{2r^2} - \frac{Z}{r}\right\}\phi = E\phi, \qquad (1)$$

---

[1] A shorter version of this note is published in **Physics Today**, Readers' Forum, **8**, p.8, 2014.

where Z is the nuclear charge and $l$ is the angular momentum.

Obviously, this equation has no sense for $D=1$. Indeed, in one dimension a finite angular momentum requires infinite speed of a rotating particle. Therefore for D=1 the correct equation is

$$\left(-\frac{1}{2}\frac{d^2}{dx^2}-\frac{Z}{r}\right)\phi = E\phi, \qquad (2)$$

that does not follow from (1) at D=1.

But most important is that the equation (1) implies independence of the Coulomb potential upon D. This could be considered as correct before it became clear that the Coulomb law follows from Maxwell equations, when one considers a field generated by a point-like electric charge. So, to obtain the Coulomb law for a two-dimensional space, one has to consider Maxwell equations in a two-dimensional world. This equation for the considered case looks as

$$-\left(\frac{\partial^2}{\partial x^2}+\frac{\partial^2}{\partial y^2}\right)\varphi^{(2)}(r) = Z\delta^{(2)}(r), \qquad (3)$$

where $\delta^{(2)}(r)$ is the two-dimensional delta-function and $r^2 = x^2 + y^2$.

This leads to

$$\varphi^{(2)}(r) = -2Z\ln(r/r_0) \qquad (4)$$

instead of $\varphi^{(3)}(r) = Z/r$ $r$-dependence in the three dimensional case. Here $r_0$ is a cutoff length that depends upon the problem at hand.

If one takes into account the D-dependence of the Coulomb potential, derivations performed in the considered paper became meaningless. Indeed, by using an unrealistic equation (1), how one can believe that it illuminates quite realistic and physical postulates of Bohr?!

It is known since long ago a number of papers that consider two-dimensional Hydrogen atom (see e.g. [6]) that use for this or that reason the incorrect for the two-dimensional case potential $Z/r$ instead of the correct one (4). It would be interesting to perform calculations with (4), instead of $Z/r$.

It would be of interest to find for completeness the shape of a "Coulomb" potential in the case of any arbitrary dimension D>3. This could be achieved by generalizing (3) in a quite natural way

$$-\sum_{i=1}^{i=D}\frac{\partial^2}{\partial x_i^2}\varphi^{(2)}(r) = Z\delta^{(2)}(r). \qquad (3)$$

Here $r^2 = \sum_{i=1}^{i=D} x_i^2$.

This derivation would also permit to check whether indeed at $D\to\infty$ one arrives to a simple classical picture. I doubt.

Quite interesting would be an attempt to model the transition from $Z/r$ to (4) by compressing an electron-proton-like system by analog of flat conducting mirrors. Perhaps that could be achieved in cavities of rectangular shape.

There exists another case, when the Coulomb-like potential is of interest. I mean Newton's gravitation law. Its analogue for two-dimension world is definitely of at least theoretical interest. To do this, one has to use the equation of Einstein's general relativity and to apply them for a massive body in the field of point-like source of gravitational field.